\documentclass[times,twocolumn,final,nopreprintline]{elsarticle}

\usepackage{framed,multirow}
\usepackage{rotating}
\usepackage{tikz}
\usepackage{wrapfig}
\usepackage{lscape}
\usepackage{rotating}
\usepackage{epstopdf}
\usepackage{commath}

\usepackage{amssymb,amssymb,amsfonts}
\usepackage{latexsym}
\usepackage{algorithmic}
\usepackage{graphicx}
\usepackage{caption}
\usepackage{subcaption}
\usepackage{textcomp}
\usepackage{caption}
\usepackage{multirow}
\usepackage{float}
\usepackage{breqn}
\usepackage{subcaption}
\usepackage{xcolor}
\usepackage{adjustbox}
\usepackage{amsmath}
\usepackage{array}
\usepackage{multirow}
\usepackage{makecell}

\newcolumntype{P}[1]{>{\centering\arraybackslash}p{#1}}

\usepackage{url}
\usepackage{xcolor}

\definecolor{newcolor}{rgb}{.8,.349,.1}

\begin{document}

\title{SynCLay : Interactive Synthesis of Histology Images from Bespoke Cellular Layouts}%

\author[1]{Srijay Deshpande\corref{cor1}}
\ead{srijay.deshpande@warwick.ac.uk}
\cortext[cor1]{Corresponding author at Department of Computer Science, University of Warwick, UK}
\author[1]{Muhammad Dawood}
\author[1]{Fayyaz Minhas}
\author[1,2,3,4]{Nasir Rajpoot\corref{cor1}}
\ead{n.m.rajpoot@warwick.ac.uk}

\address[1]{Tissue Image Analytics Centre, Department of Computer Science, University of Warwick, UK}
\address[2]{The Alan Turing Institute, London, UK}
\address[3]{Department of Pathology, University Hospitals Coventry \& Warwickshire, UK}
\address[4]{Histofy Ltd, Birmingham, UK}

\begin{abstract}

 Automated synthesis of histology images has several potential applications in computational pathology. However, no existing method can generate realistic tissue images with a bespoke cellular layout or user-defined histology parameters. In this work, we propose a novel framework called SynCLay (Synthesis from Cellular Layouts) that can construct realistic and high-quality histology images from user-defined cellular layouts along with annotated cellular boundaries. Tissue image generation based on bespoke cellular layouts through the proposed framework allows users to generate different histological patterns from arbitrary topological arrangement of different types of cells (e.g., neutrophils, lymphocytes, epithelial cells and others). SynCLay generated synthetic images can be helpful in studying the role of different types of cells present in the tumor microenvironmet. Additionally, they can assist in balancing the distribution of cellular counts in tissue images for designing accurate cellular composition predictors by minimizing the effects of data imbalance. We train SynCLay in an adversarial manner and integrate  a nuclear segmentation and classification model in its training to refine nuclear structures and generate nuclear masks in conjunction with synthetic images. During inference, we combine the model with another parametric model for generating colon images and associated cellular counts as annotations given the grade of differentiation and cellularities (cell densities) of different cells. We assess the generated images quantitatively using the Frechet Inception Distance and report on feedback from trained pathologists who assigned realism scores to a set of images generated by the framework. The average realism score across all pathologists for synthetic images was as high as that for the real images. We demonstrate that the proposed framework can be used to add new cells to a tissue images and alter cellular positions. We also show that augmenting limited real data with the synthetic data generated by our framework can significantly boost prediction performance of the cellular composition prediction task. The implementation for the proposed SynCLay framework is available at \url{https://github.com/Srijay-lab/SynCLay-Framework}
 
\end{abstract}

\maketitle


\section{Introduction}

Generative modeling of histology images has been an active area of research in the field of computational pathology (CPath) in recent years. A number of approaches have been presented for generating high-quality synthetic tissue images (\citep{quiros2019pathology, fmahmooddeepsegmentation, safron}. Synthetic histology images are useful in development of algorithms for downstream predictive modeling tasks \citep{robustlabelsynthesize, Levine2020SynthesisOD, safron}. Moreover, they can also be utilized in education \citep{educationapplication}, clinical quality assurance \citep{qualityassurance}, for addressing privacy issues, and for overcoming ethical and legal barriers related to sharing image data \citep{privacyapplication, jacobkathermsi}. 

Most of the aforementioned use cases of synthetic histology image generation require image synthesis from custom cellular layouts or user-defined tissue parameters such as cellular composition or disease grade. Another key requirement for such models is that they be able to generate nuclear or tissue level annotations along with a generated image in case these generated images are to  be used for training deep learning models for cellular composition prediction or nuclear segmentation where expert annotations can be difficult and time consuming to obtain. 

Despite a recent increase in synthetic histology image generation methods, there are no existing methods that can generate  images and their associated annotations conditioned on input cellular layouts or user-defined histology parameters. The term cellular layout refers to a two-dimensional plane where a user can arrange arbitrary types of cells at different spatial locations on the plane. Such a bespoke layout provides flexible control over nuclei in the generated image which can then be used for various applications. For instance, they can potentially be used to overcome the data imbalance problem for CPath tasks such as cellular composition prediction in which the number of certain types of nuclei such as neutrophils can be quite small in comparison to other types such as epithelial or connective cells. 

In this paper, we propose a novel generative framework called SynCLay (\textit{Synthesis from Cellular Layouts}) for generating synthetic histology images from bespoke cellular layouts. To the best of our knowledge, this is the first method that can generate annotated histology images from bespoke cellular layouts. The proposed framework can also be used to generate tissue images from a set of user-defined parameters such as grade of cancer differentiation and proportions of different types of cells in an image. For this purpose, we use SynCLay in conjunction with a parametric model to first generate a cellular layout from user-defined parameters \citep{kovacheva2016model}. This integration also allows construction of visually realistic complex multi-cellular structures like glands. In order to improve the visual quality of generated nuclei as well as generating nuclear masks alongside tissue images, the proposed framework proposes a novel integration of a nuclear segmentation and classification model called HoVerNet \citep{graham2019hover}. 

We demonstrate the significance of bespoke synthetic images generated using the proposed approach in balancing the training data for training cellular composition prediction and nuclei presence detection algorithms for which training data can be highly biased due to data imbalance across different types of nuclei. We show that by artificially increasing the counts of minority class cells, synthetic images can be used to minimize data imbalance in training data and thereby improve the performance of cellular composition prediction algorithms for rarely occurring cell types and also detecting their occurrence in images. We also present a detailed quantitative comparison with other state-of-the-art methods for histology image synthesis based on the Frechet Inception Distance \citep{heusel2017gans} metric.  Major contributions of this paper are listed below:

\begin{enumerate}

\item We propose an interactive framework SynCLay that can generate tissue images from bespoke cellular layouts. The proposed framework also allows the user to generate custom tissue images by adding and moving cells by changing the cellular layout. 

\item The proposed approach also allows generation of realistic synthetic histological images and their associated cellular counts from a set of user-defined parameters such as grade of differentiation of cancer and cellularities (cell densities) of different kind of cells.

\item We incorporate a nuclear morphology loss function based on a nuclear segmentation model (called HoverNet~\citep{graham2019hover}) into the framework to improve the quality of generated nuclei. This integration also enables the simultaneous generation of nuclei segmentation masks alongside images.

\item We assess the realism of synthetic images with the help of 4 trained pathologists and show that the quality of generated images is comparable with their real counterparts. 

\item Finally, we highlight the benefit of using synthetically generated colorectal histology image data using the proposed framework for downstream cellular composition prediction and nuclei presence detection tasks.

\end{enumerate}

\begin{figure*}
\centering
\begin{subfigure} {\columnwidth}
 \makebox[\textwidth]{\includegraphics[width=380pt,height=320pt]{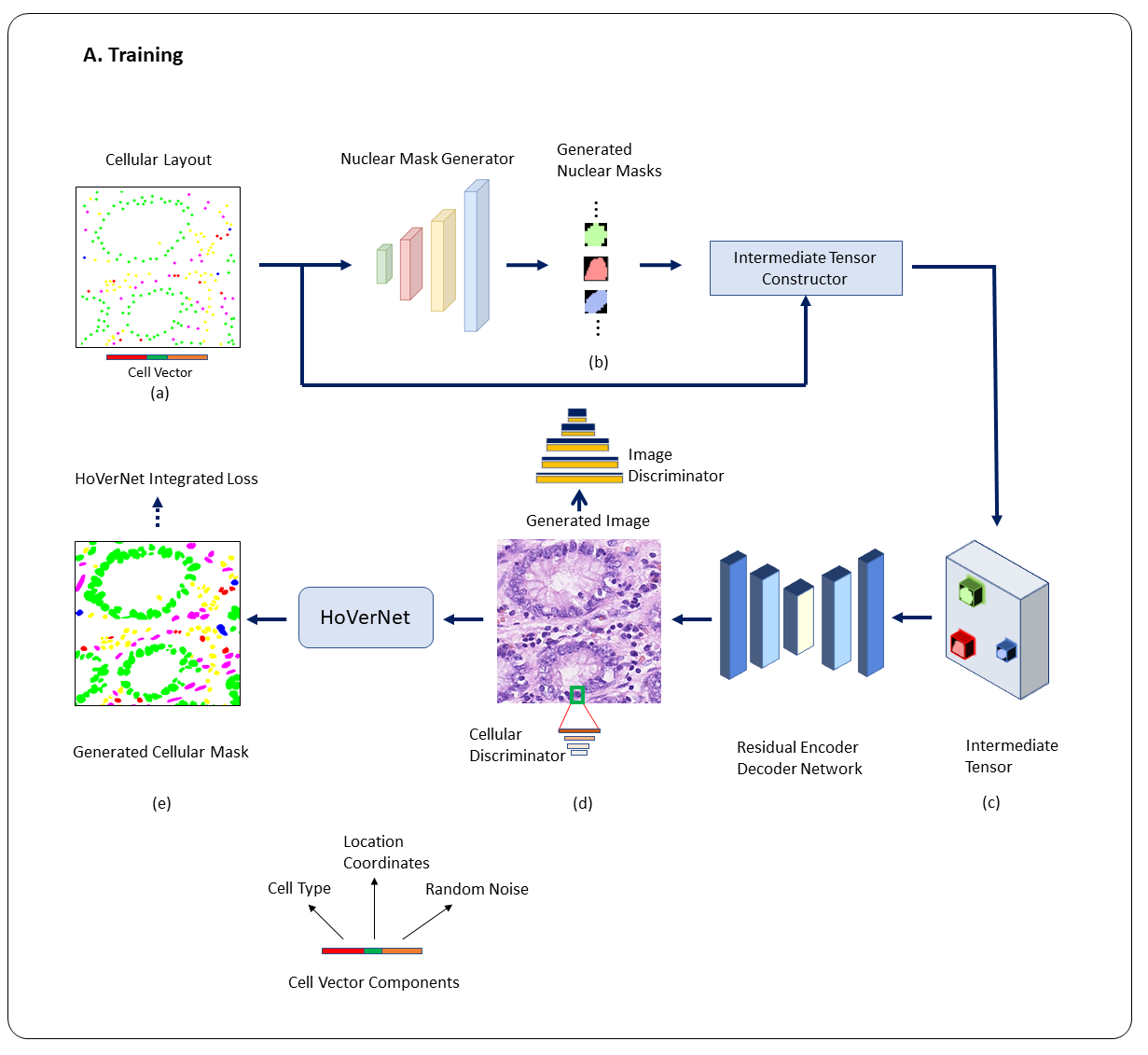}}
\end{subfigure}

\vspace{15pt}
\centering
\begin{subfigure} {\columnwidth}
 \makebox[\textwidth]{\includegraphics[width=420pt,height=90pt]{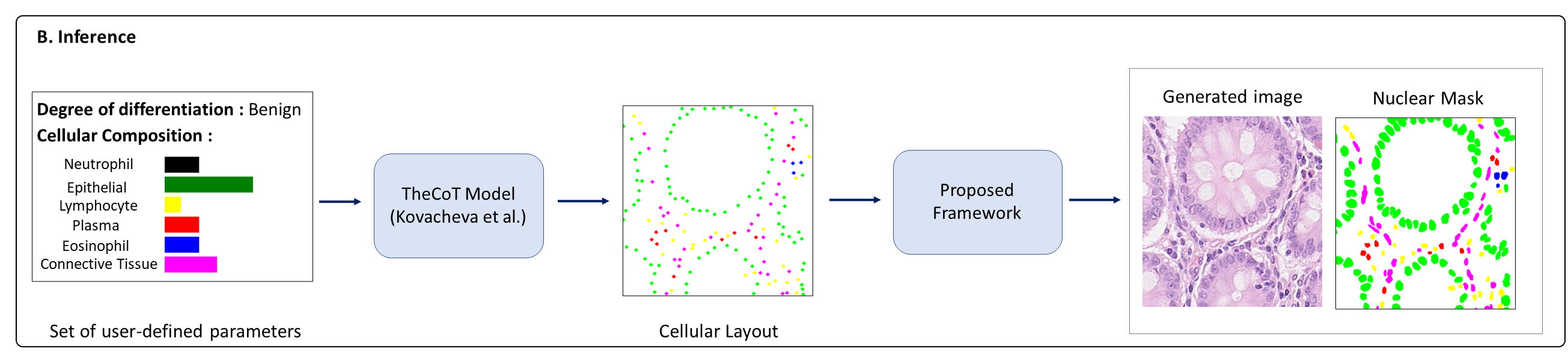}}
\end{subfigure}
\caption{An overview of the proposed SynCLay  framework for generating colon histology images from the bespoke cellular layouts. The cellular layout (a)  define the spatial location of different types of cells (shown by different colors). Each cell is represented by a cellular vector. Each cellular vector is passed through the mask generator neural network generating binary cellular masks (b). The bounding box for each cell is computed by its spatial location and its size. The size can be defined as a pair of lengths of horizontal and vertical diameters of a cell. These size coordinates are either taken directly from the datasets or constructed using the statistics of cellular sizes from the available data. The cellular vectors are then element-wise multiplied with the constructed binary cellular masks generating masked embeddings which are then wrapped into the positions of bounding box coordinates using the bilinear interpolation algorithm giving an intermediate tensor (c). The intermediate tensor is passed through the residual encoder decoder neural network generating the colon histology image (d). We integrate the HoVer-Net into the framework which enables the generation of the nuclei segmentation mask (e) by passing the generated tissue image to the HoVer-Net. The HoVer-Net integrated loss component also assists in curating the cellular structures inside the generated tissue image. We combine the proposed framework with the existing TheCoT framework while inference. The TheCoT model takes user-defined parameters like grade of differentiation of cancer and cellularities cells and construct the cellular layout which mimic the actual location structure of nuclei. The cellularities are real numbers between 0 and 1 are shown in different colors for respective nuclei types in the leftmost of inference part. The cellular layout is then passed to the proposed framework that generates the pair of a histology image and its nuclei segmentation mask.}
\label{concept_figure}
\end{figure*}

The remainder of this paper is organized as follows. In the subsequent section, we go through the related work on relevant methods of image generation in CPath. In section {\color{red}{\ref{materials_methods}}}, we provide details of various components in the proposed framework. In Section {\color{red}{\ref{exp_results}}}, we present results demonstrating the efficacy of the proposed framework. We evaluate the importance of architecture design of SynCLay with various ablation experiments, followed by discussion in {\color{red}{\ref{discussion}}}. Finally, we conclude with future directions.

\section{Related Work}

In the last decade, wide adoption of Generative Adversarial Networks or GANs \citep{goodfellow2014generative} has led to the generation of realistic and high-quality synthetic tissue images in CPath. For instance, \cite{quiros2019pathology} presented Pathology-GAN to generate high-fidelity cancer tissue images. The model also learned pathologically meaningful representations within cancer tissue images that allowed it to perform linear arithmetic operations to change high-level tissue image characteristics. \cite{Levine2020SynthesisOD} presented an adversarial learning approach based on ProGANs to generate high-quality histology images of size $1024\times1024$ pixels. Though these approaches were able to generate high-fidelity histology images, they didn't produce matching annotations which are required for development of machine learning algorithms for various tasks in computational pathology. 

Some researchers have investigated generating synthetic pathology images conditioned on tissue component masks \citep{mirza2014conditional, senaras2018creating}. \cite{robustlabelsynthesize} proposed an unsupervised pipeline to construct both histopathology tissue images and their corresponding  nuclei masks to train a supervised nuclear segmentation algorithm. Their image generation model used the output from its discriminator network to assign instance weights for training the nuclear segmentation algorithm. The success of conditional GANs \citep{mirza2014conditional} (cGANs) in generating high-fidelity images conditioned on known ground truth inspired researchers to adapt them for tissue image synthesis. \cite{senaras2018optimized} proposed a cGAN based model to construct breast cancer tissue images conditioned on input nuclear masks. As the generative models are sensitive to artifacts in synthetic images, \cite{simgan} proposed an unsupervised approach to add realism in generated images and stabilize GAN training. These methods either assume input component masks are already present, or require explicit construction of component masks by generating random shapes of respective tissue components like nuclei, which can be erroneous and may not be realistic. Moreover, this process of crafting component masks can be tricky for various nuclear structures. Generating synthetic images along with component masks simultaneously is therefore desirable as it potentially reduces the cost of annotations and also constructs realistic annotated pairs. Furthermore, none of these methods work on placing nuclei in a proper structure i.e., placing epithelial cells around the glands. It may result in unrealistic nature of nuclei locations.

\section{The Proposed Method}
\label{materials_methods}

Our aim is to develop an interactive framework that enables generation of colon tissue images from user-defined cellular layouts. The cellular layout can be described as a plane where users can arrange cells of distinct types on its 2-d spatial locations as shown in Figure \ref{concept_figure}. We feed this layout to the proposed framework which models spatial dependencies between cells and constructs histology image of size $256\times256$ pixels through a series of different neural networks. Tissue image generation from the cellular layout allows a user to control the locations and types of cells in the colon histology image. We integrate the framework with pretrained HoVer-Net \citep{graham2019hover} which allows generation of nuclei segmentation masks simultaneously with images. The proposed framework also allows generation of histological images by using cellular composition (counts of different types of cells) as input. This is achieved by generating a cellular layout from the input cellular composition vector through the  TheCOT framework \cite{kovacheva2016model}. Thus, the entire framework has two major components: First, we construct a cellular layout using from a set of user-defined parameters such as cellularities of different cells and grade of differentiation of cancer. Second, the proposed framework takes the cellular layout as an input and generates tissue images along with their nuclei masks. An overview of the framework is given in Figure \ref{concept_figure}. The implementation can be found here\footnote{https://github.com/Srijay-lab/SynCLay-Framework}. Below we describe the main components of the proposed framework:

\subsection{Cellular Layout Representation}

The input cellular layout can be assumed as a set of $n$ cells of different types and the corresponding locations on two dimensional plane. Each cell, indexed by $k = 1\ldots n$, is characterized by a triplet $c_k = (t_k,l_k,z_k)$ comprising a one hot encoding vector $t_k$ to specify nucleus or cell type, a two dimensional location vector $l_k$ and random noise sampled from the Gaussian noise $z_k \sim \mathcal{N}(0,1)$, $|{z_k}| = 4$. The components of the location vector $l_k$ are normalized to the range $[0,1]$.  The Gaussian noise is added into the cell vectors in order to ensure variable appearance of generated cellular objects in the final image. 

\subsection{Generation of Binary Cellular Masks}
\label{generationbinarycellular}

The cellular vectors $\{c_k \mid k = 1,2,..., n \}$ are input to a \textit{mask generator network} $M$ to generate the corresponding individual cellular binary masks $\{ m_k \mid k = 1,2,...,n \}$, each of the size $64 \times 64$ pixels i.e., $m_k = M(c_k;\theta_{M})$, where $\theta_{M}$ denotes the trainable parameters of the model. The mask generator network is comprised of series of blocks having transpose convolution layer followed by the ReLU activation. The detailed architecture is given in the Appendix. 

In order to generate the histology image, we need to move to from the input cellular layout to the image domain. For this purpose, we utilize the generated binary cellular masks, cellular vectors and bounding box coordinates, and construct an intermediate tensor which holds the information needed to generate the histology image using the following procedure. The bounding box coordinates $\{ b_k \mid k = 1,2,...,n \}$ are computed from the input location parameters and the horizontal and vertical sizes of the cells. These size coordinates are either taken directly from the training datasets or can be realized from the procedure described in Section \ref{exp_results}. Each cellular vector $v_k$ is multiplied element-wise with the individual cellular mask $m_k$ to give masked embedding of size $8 \times 64 \times 64$ which is then wrapped to the position of bounding box using a fixed bilinear interpolation function $F$ (\cite{bilinearinterpol}). This gives an intermediate tensor of dimensionality $8 \times 256 \times 256$, i.e., $T = F(\{v_k,m_k,b_k \mid k = 1,2...n\})$, that holds information about all cells in the given input layout. 

\vspace{-3mm}

\subsection{Histology Image Generation}

\vspace{-1mm}

After generating the intermediate tensor $T$, we feed it to the encoder-decoder residual network \citep{scene_graph_2}. The image-to-image translation encoder-decoder network is used as an image generator to construct the final tissue image $I = E(T, \theta_E)$ of size $256 \times 256$ pixels. The network consists of a series of residual blocks which transform the input tensor into an output image. The exact architecture of the encoder-decoder residual network is provided in the Appendix. 

\vspace{-3mm}

\subsection{HoVer-Net Integration}

In order to support generation of nuclear masks and to refine the quality of generated nuclei, we incorporate a nuclear segmentation and classification algorithm called HoVer-Net \citep{graham2019hover} into our framework. We pass a generated image through a pretrained HoVer-Net model denoted by $H$ and compute the nuclei mask $Y$ as shown in Figure \ref{concept_figure} i.e., $Y = H(I)$. The pretrained HoVer-Net is kept frozen after incorporating into our framework. Therefore, it does not have any trainable parameters. This setting allows HoVer-Net integrated loss to get incorporated into the loss function for training the framework.

\subsection{Discriminators}
\label{discriminators_section}

\vspace{-1mm}

We employ two discriminator neural networks to make generated tissue image and its cellular components appear realistic: \textit{image discriminator} $D_I(I;\theta_I)$ for the generated tissue image $I$, and the \textit{cellular discriminator} $D_C(I_{c_i};\theta_C)$ for cellular components $\{I_{c_i} \mid i=1,2..c\}$ inside the tissue image, where $c$ is the number of cells within it; $\theta_I$ and $\theta_C$ denote the respective trainable parameters of those discriminators. The first  discriminators employ the same architecture as the PatchGAN (\cite{pix2pix}) discriminator which predicts the realism of the different portions from the generated component mask and the tissue image, respectively. The adversarial losses based on these discriminators ensure tissue component masks and tissue images appear realistic. 

The architecture of the \textit{cellular discriminator} is comprised of a series of convolution operations and predicts a single score of realism for the generated glandular portions cropped out from the final tissue image based on input bounding boxes, and resized to a fixed size using bilinear interpolation (\cite{bilinearinterpol}). It ensures the generated cells, one by one, appear real with their micro-components like nuclei and cytoplasm.

\subsection{Loss Function Terms}
\label{loss_components}

Here we give details about all loss terms used in our framework. The training loss of the proposed framework is composed of several terms as it involves multiple networks. as described below:

\noindent \textbf{Cellular masks reconstruction loss:} This component penalize the difference between ground truth $\{ \hat{m_k} \mid k = 1,2...n \}$ and generated individual binary glandular masks $\{ m_k \mid k = 1,2...n \}$ using the mean square error (MSE) as follows,

  \begin{equation}
      L_M(\hat{m_k},m_k ; \theta_M) =  \sum_{k=1}^n MSE(\hat{m_k},m_k)
  \end{equation}

where $\hat{m_k}$ is the ground truth, $m_k$ is the generated binary cellular mask generated in \ref{generationbinarycellular}, and $n$ is the number of cells in the tissue image. As we saw in section \ref{materials_methods}, $m_k$ is dependent on the trainable parameter $\theta_M$.\\

\noindent \textbf{Image Reconstruction Loss:} This term captures the reconstruction error between ground truth $\hat{I}$ and generated tissue image $I$ using the $L1$ difference,

  \begin{equation}
      L_I(\hat{I}, I ; \theta_M, \theta_E) =  \| \hat{I} - I\|_1
  \end{equation}
where $\hat{I}$ is ground truth and $I$ is the generated tissue image. \\

\noindent \textbf{HoVer-Net Integration Loss:} This term captures the label prediction error between ground truth nuclei segmentation mask $\hat{Y}$ and HoVer-Net predicted nuclei segmentation mask $Y$ using the cross entropy loss function. For this purpose, we use pretrained HoVer-Net and freeze its model parameters in order to make generated tissue image aligned with nuclei segmentation mask as a function of HoVer-Net model. 

  \begin{equation}
      L_H(\hat{Y}, Y ; \theta_M, \theta_E) =  CrossEntropy(\hat{Y}, Y)
  \end{equation}

\noindent \textbf{Adversarial Loss Terms}: We employ an adversarial loss function (\cite{goodfellow2014generative}) for both discriminators used in the SynCLay framework. A discriminator $D_t(X;\theta_t)$ attempts to maximize the loss by classifying the input image $X$ generated by the generator function $G(X;\theta_G)$ which tries to minimize it, where $t$ denotes the type of the discriminator among the \textit{image discriminator} ($t=I$) and \textit{cellular discriminator} ($t=C$), $\theta_t$ and $\theta_G$ denotes the set of trainable parameters of the respective networks. The adversarial min-max loss function is given by,

\vspace{-5mm}

\begin{equation}
\begin{aligned}
\min_{\theta_G} \max_{\theta_t} L_{GAN}^t(X; \theta_G,\theta_t) = E_{X\sim p_{data}(X)}[log D_{t}(X;\theta_t)] + \\ E_{X \sim p_{X}(X)}[log(1-D_{t}(G(X;\theta_G);\theta_t)]
\end{aligned}
\label{advloss}
\end{equation}

Therefore, the two adversarial loss terms: $L_{GAN}^I$ and $L_{GAN}^C$ for tissue image and the cellular components cropped out from the tissue image, respectively, their expressions can be realized by putting $t=I$ and $t=C$ in Equation (\ref{advloss}), 

\begin{equation}
\begin{aligned}
L_{GAN}^I(I ; \theta_M, \theta_E, \theta_I ) = \\ \min_{\theta_G} \max_{\theta_I} L_{GAN}^I(I; \theta_G = \{\theta_M, \theta_E\},\theta_I)
\end{aligned}
\label{advloss_image}
\end{equation}

\begin{equation}
\begin{aligned}
L_{GAN}^C(\{I_{c_i}\} ; \theta_M, \theta_E, \theta_C ) = \\ \sum_{i=1}^c \min_{\theta_G} \max_{\theta_C} L_{GAN}^I(I_{c_i}; \theta_G = \{\theta_M, \theta_E\},\theta_C)
\end{aligned}
\label{advloss_image}
\end{equation}

The overall learning problem can be cast as a the adversarial optimization problem based on the linear combination of adversarial and reconstruction losses. The framework is trained by minimizing the following objective $\mathcal{L}$:

\begin{dmath}
\centering
\mathcal{L} = \lambda_1 L_I + \lambda_2 L_M + \lambda_3 L_H + \lambda_4 L_{GAN}^I + \lambda_5 L_{GAN}^C
\label{losseq}
\end{dmath}

where $\lambda_1, \lambda_2, \lambda_3$, $\lambda_4$ and $\lambda_5$ denote the weights of corresponding loss components.

\subsection{Inference using User-defined Parameters}

The trained framework is able to generate histology images from input cellular layouts. The cells and their locations can be altered in order to change the appearance of generated images. The sample can be seen in Figure \ref{alter_layout}. To give flexibility of generating annotated images, we utilize a parametric model to construct the cellular layout from user-defined parameters.

To enable this, we integrate the proposed framework with the existing TheCoT model \citep{kovacheva2016model} in order to enable the image generation from user-defined parameters such as grade of differentiation of cancer and cellularities (cell densities) of different types of cells. The rest of the parameters consumed by the TheCoT model such as image size = $256\times256$, magnification = 40$\times$, cell overlap = 0 are kept fixed in this experiment.

The framework first estimates the number of cells and glands based on the image size and magnification. It then computes number of each of the distinct cells based on their input cellularities between 0 and 1. Based on the grade of differentiation parameter, it draws glands and place epithelial cells along its surface. The rest of the cells are placed according to uniform distribution. The sample cellular layout generated using TheCoT model can be seen in Figure \ref{concept_figure}.

\section{Experiments and Results}
\label{exp_results}

In this section, we present visual and quantitative results of the quality of generated images and comparison of SynCLay with other state-of-the-art models employed for high-quality image generation. We also assess the quality of synthetic images with the help of expert pathologists. We demonstrate the utility of bespoke synthetic images in reducing the class imbalance in training datasets and for improving the performance of the cellular composition prediction algorithm. Finally, we demonstrate the importance of the SynCLay framework design and validate its loss components using Ablation study. As a side experiment, we also investigate the results of employing graph convolution neural network to model spatial dependencies between cells on the cellular layout for generation of synthetic images.

\begin{figure*}[]
\centering
\includegraphics[width=380pt,height=700pt,keepaspectratio]{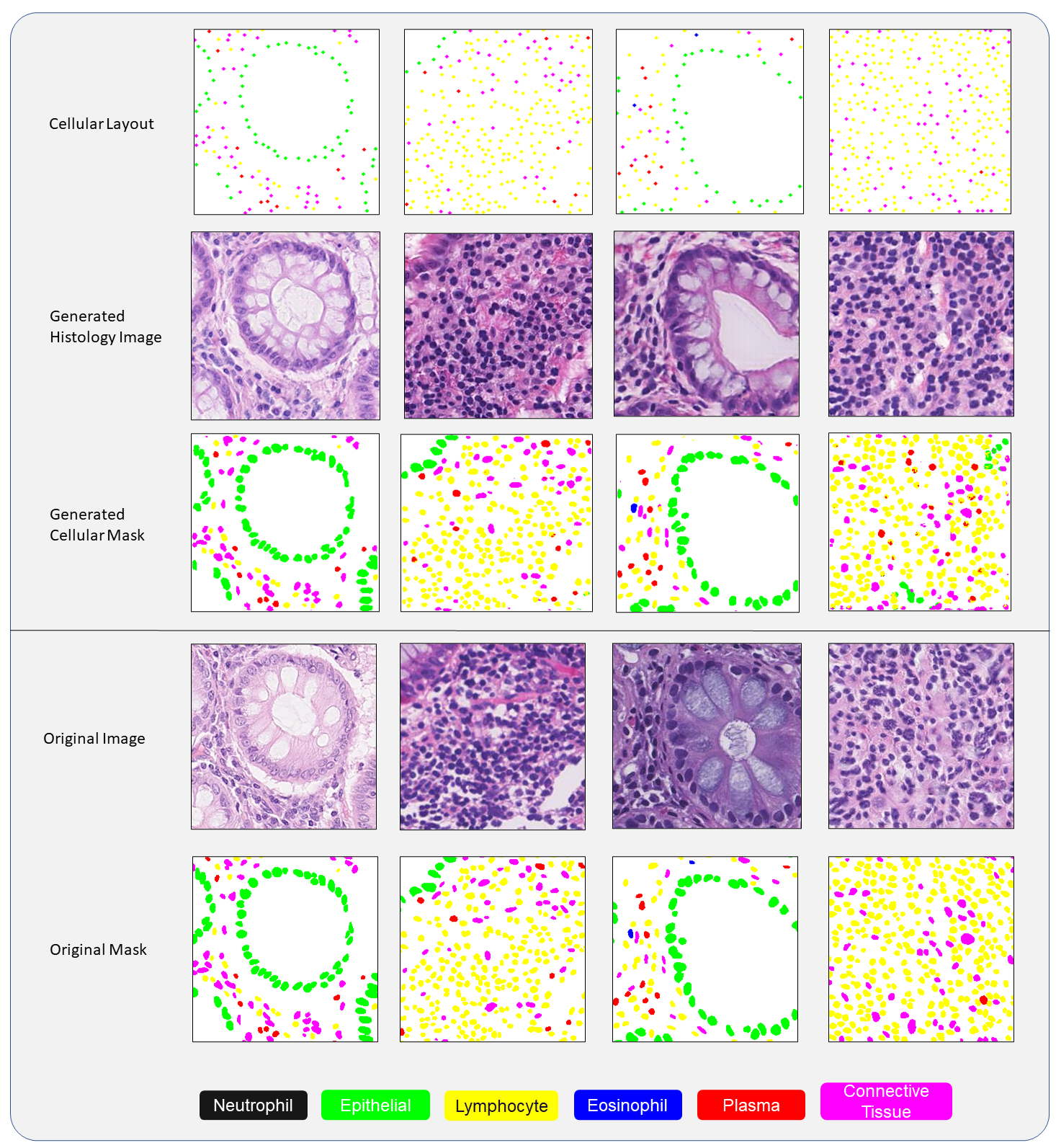}
\caption{Samples of generated images and nuclei segmentation masks from the CoNiC dataset. The locations of distinct cells are shown in different colors in the topmost row of cellular layouts.}
\label{results_1}
\end{figure*}

\begin{figure*}[]
\centering
\includegraphics[width=380pt,height=700pt,keepaspectratio]{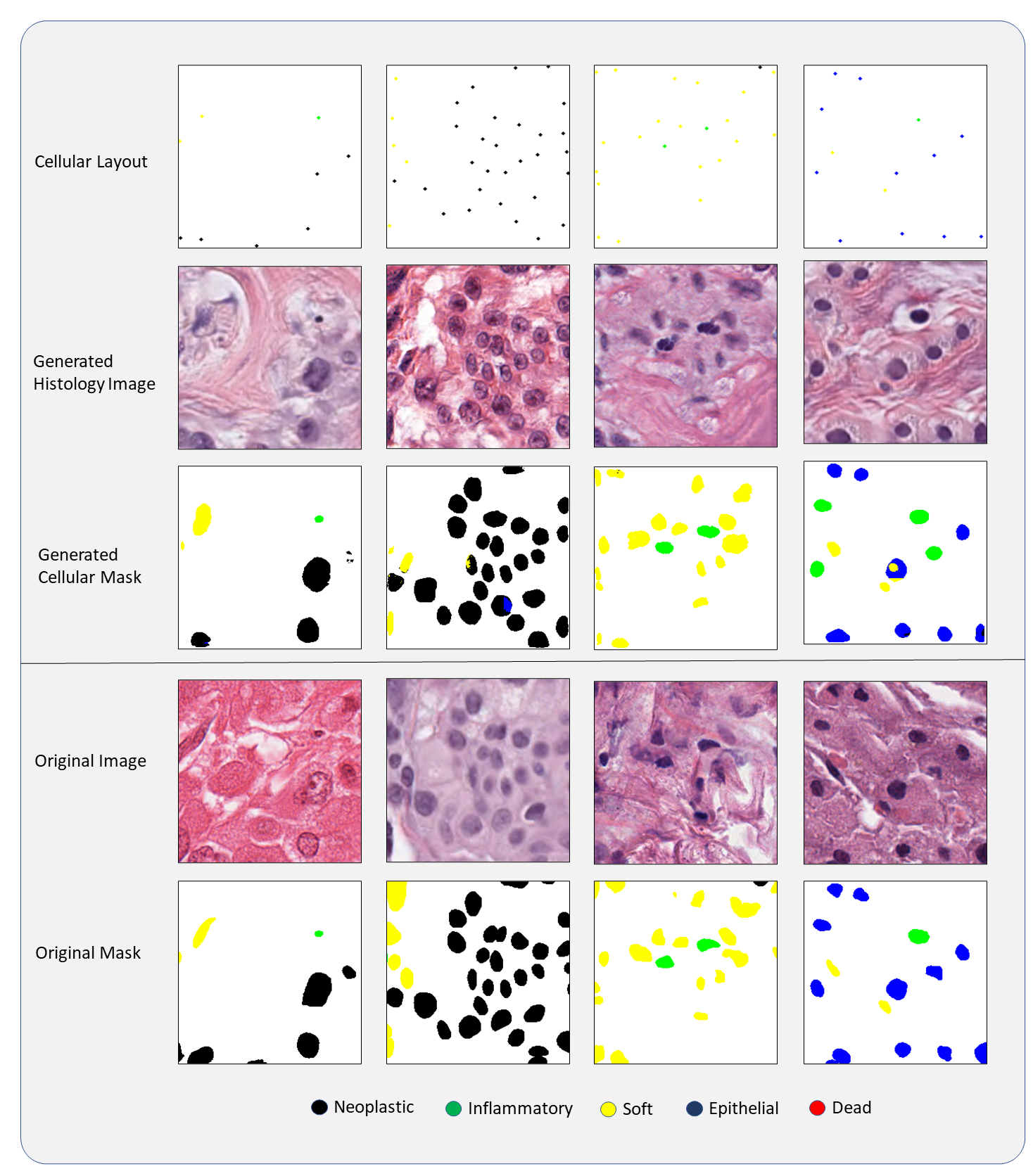}
\caption{Samples of generated images and nuclei segmentation masks from the PanNuke dataset. The locations of distinct cells are shown in different colors in the topmost row of cellular layouts.}
\label{results_2}
\end{figure*}

\begin{figure*}[]
\centering
\includegraphics[width=\textwidth,height=\textheight,keepaspectratio]{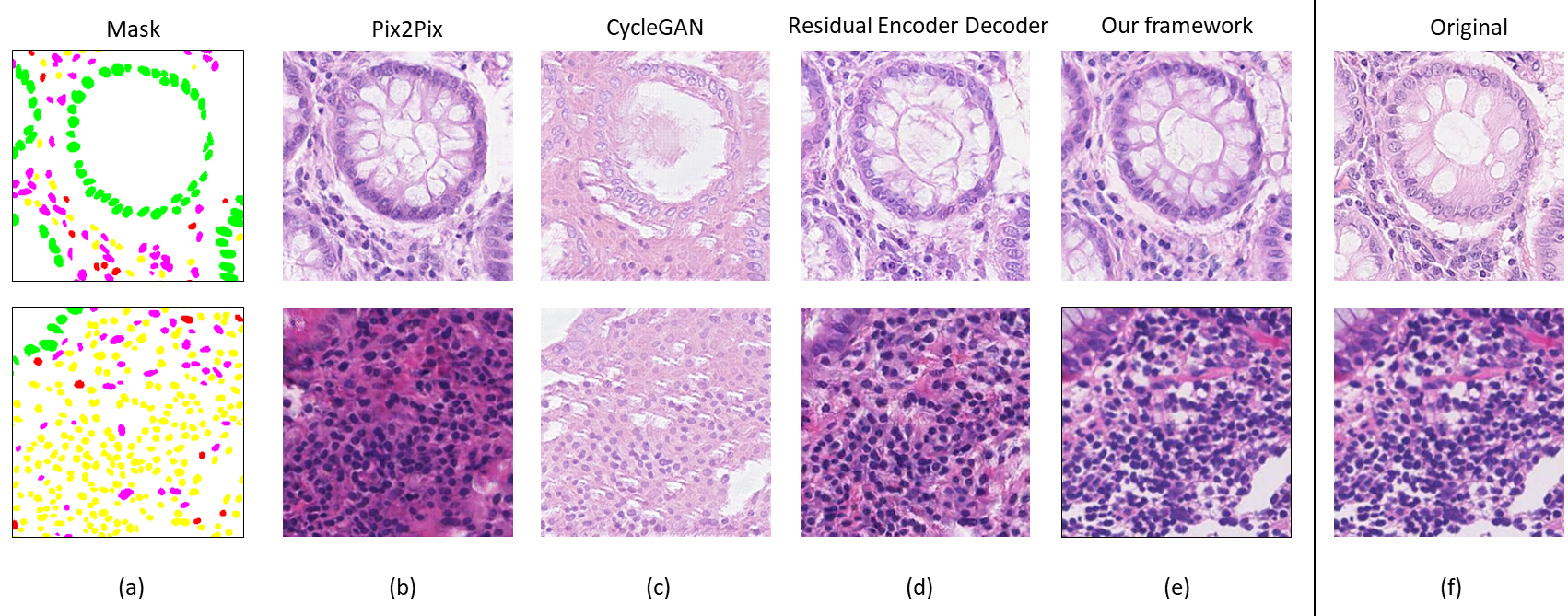}
\caption{From left to right: (a) Input nuclei mask for conditional generative models used for comparison, (b) Pix2pix \cite{pix2pix}, (c) CycleGAN (\cite{cyclegan}), (d) Residual Encoder Decoder Network (e) Proposed framework, (f) Original image}
\label{results_3}
\end{figure*}

\subsection{Datasets}
\label{datasets_section}

For training and performance evaluation of the proposed SynCLay framework, we require the image data annotated with cellular layouts. In this work, we consider two datasets for our experiments: CONIC\footnote{\label{conicurl}https://conic-challenge.grand-challenge.org/} \citep{conic1,conic2} and PanNuke\footnote{https://jgamper.github.io/PanNukeDataset/} \citep{pannuke1,pannuke2}.

\noindent \textbf{CoNiC Dataset}: This dataset is collected from the CoNiC challenge \citep{conic1,conic2}. Overall, it contains 4,981 Haematoxylin and Eosin stained colon histology image of size $256\times256$ coupled with corresponding nuclei segmentation mask of the following nuclei types: epithelial, lymphocyte, plasma, eosinophil, neutrophil and connective tissues. From these images, we use 3918 images for training (CoNiC train set) and rest for testing (CoNiC test set). 

\noindent \textbf{PanNuke Dataset}: This dataset includes semi-automatically generated nuclei instance segmentation masks with exhaustive nuclei labels across 19 different tissue types. The dataset consists of 481 visual fields, of which 312 are randomly sampled from more than 20K whole slide images at different magnifications, from multiple data sources. From these regions, we keep 2656 images for training (PanNuke train set) and 2522 for testing (PanNuke test set). 

Both of these datasets contain histology images and their corresponding nuclei masks. In order to obtain cellular layouts, we adopt the following procedure: First, we extract nuclei objects from the nuclei segmentation masks using OpenCV (Open Source Computer Vision Library) python library (\cite{opencv}), and locate their centroids. Then we collect bounding boxes for each of the cells using boundingRect() function of the same library.

\begin{figure*}[hbt!]
\centering
\includegraphics[width=\textwidth]{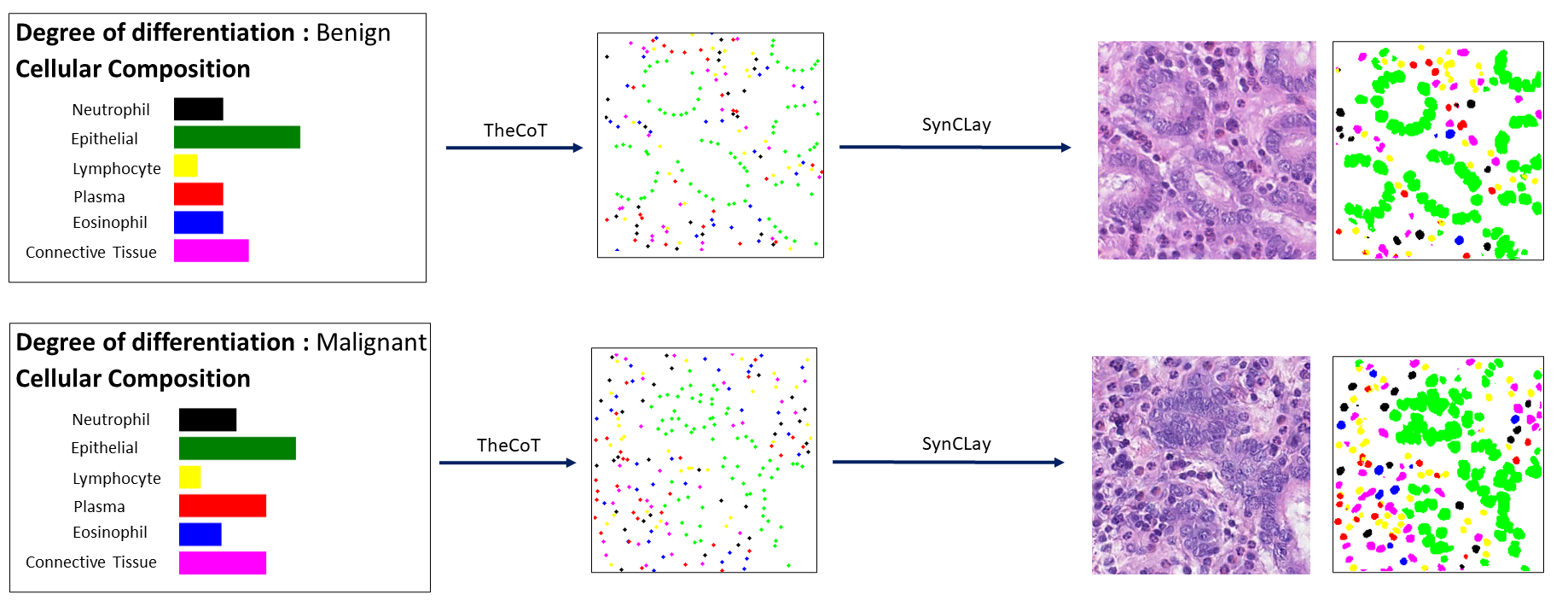}
\caption{Sample generations of colon cancer images given the cellularities (cell densities) of cells in (0,1) range. The input layouts are constructed with the help of TheCoT model \citep{kovacheva2016model}. Later, the SynCLay  model generates the tissue image and its associated nuclei segmentation mask.}
\label{thecot_generated}
\end{figure*}

\begin{figure*}[hbt!]
\centering
\includegraphics[width=340pt,height=680pt,keepaspectratio]{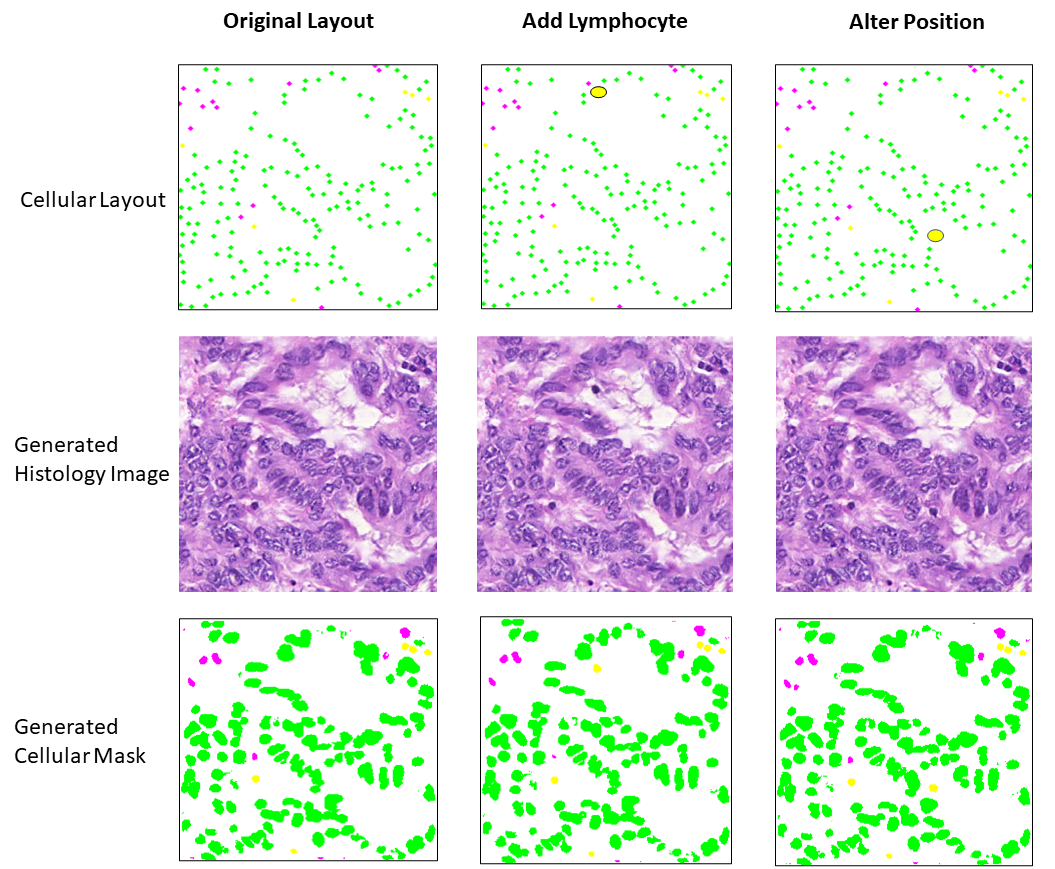}
\caption{Change in the appearance of synthetic colon tissue image and its nuclei mask after adding lymphocyte on its cellular layout. The last column shows image alterations after moving the added lymphocyte.}
\label{alter_layout}
\end{figure*}

\subsection{Model Training}

We perform the training of SynCLay framework in two phases. First, we train the framework without HoVer-Net for 100 epochs. We train HoVer-Net separately on the same training set of respective dataset. Later, we integrate this pre-trained HoVer-Net into the SynCLay and perform iterations of next 20 epochs. 

The framework is trained with Adam optimization and an initial learning rate $10^{-4}$, initial momentum 0.5 and batch size of one. While training, we set the loss components weighing coefficients: $\lambda_1=1$, $\lambda_2=1$, $\lambda_3=0.1$, $\lambda_4=0.01$, $\lambda_5=1$.

\subsection{Visual Assessment}
\label{visual_assessment}

Figures \ref{results_1} shows generated colon images from the CoNiC test set. We observe that shapes, morphological characteristics, glands, glandular lumen, and cellular appearances are preserved in the generated images, which resemble the corresponding real images quite closely. In addition, although cells can be clearly distinguished, some moderate deformities in epithelial cells are visible. We can also notice the actual nuclei segmentation mask closely resemble the output segmentation masks. Visual results on a representative image from the PanNuke test dataset are shown in Fig. \ref{results_2}. As with synthetic images from CoNiC dataset, distinct nuclei structures are apparent in generated images from the PanNuke dataset. For instance, lymphocytes that are usually in darker shades, are clearly visible in generated images and computed segmentation masks (in yellow color).

\begin{table*}[hbt!]
\centering
\begin{tabular}{|l|c|c|c|c|c|}
\hline
\textbf{Images} &
  \textit{$P_1$} &
  \textit{$P_2$} &
  \textit{$P_3$} &
  \textit{$P_4$} &
  \textbf{Mean} \\ \hline
Real Images      & 9.80 & 7.13 & 7.73 & 8.6  & 8.31 ± 1.00 \\ \hline
Synthetic Images & 9.90 & 6.70 & 6.73 & 8.33 & 7.92 ± 1.32 \\ \hline
Real Nuclei      & 9.73 & 7.93 & 7.73 & 9.33 & 8.68 ± 0.86 \\ \hline
Synthetic Nuclei & 9.80 & 7.06 & 7.4  & 8.73 & 8.25 ± 1.19 \\ \hline
\end{tabular}%
\caption{Assessment of generated tissue images and associated glands by 4 pathologists ($P_1$,$P_2$, $P_3$ and $P_4$). The images were scored between 1 (least realistic) to 10 (most realistic). The average scores show that the synthetic images are not distinguishable from the real images. The synthetic images used in this experiment are generated using SynCLay variant (with graph convolution network) discussed in the ablation study.}
\label{pathologists_assessment}
\end{table*}

In order to do visual comparison we consider state-of-the-art conditional generative models such as Pix2Pix \citep{pix2pix} and CycleGAN \citep{cyclegan} that are utilized to generate images conditioned on the input nuclei masks. We train both of these models on both CoNiC and PanNuke training sets defined in section \ref{datasets_section}. The Pix2Pix network includes the encoder-decoder generator pipeline with convolution operations for downsampling and up-convolution operations for upsampling along with skip-connections among they layers to pass low level details for the generation. CycleGAN is an image-to-image translation framework that gets trained on unpaired examples. We also consider the residual encoder-decoder network \citep{scene_graph_2} as a baseline and adapt it for the task of generating images conditioned on input masks. It is interesting to note that these existing models need input nuclei masks for inference while our model needs cellular layouts, whose construction is relatively easy and can also be done from user-defined parameters using the TheCoT model \citep{kovacheva2016model}. 

Figure \ref{results_3} shows generated images using baseline models and our models. It can be visibly noticed that the nuclei generated using our framework exhibit finer details compared to baseline models. From the figure we can notice that glandular lumen generated by our framework matches closest to that of the original. The residual encoder decoder network produces visually better tissue images compared to Pix2Pix. This also validates our choice to use residual encoder-decoder network as a backbone generator against the pix2pix generator. Figure \ref{thecot_generated} shows the image generated with the help of TheCoT framework from the set of cellularities of different types of cells. It can be observed that the generated samples appear realistic.

\begin{figure*}[hbt!]
\centering
\includegraphics[width=\textwidth, keepaspectratio]{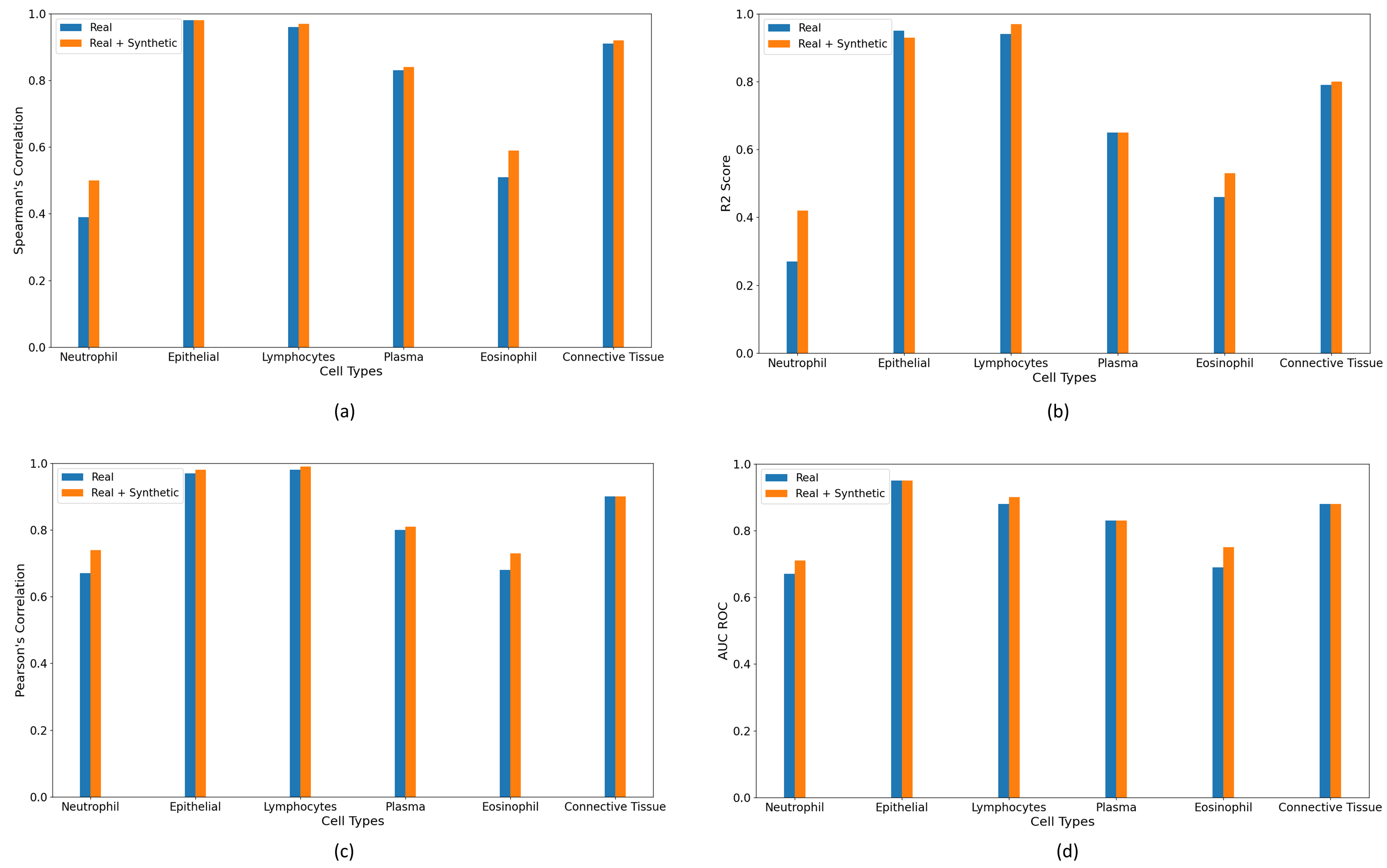}
\caption{Improvement in various performance measures, (a) Spearman's Correlation (b) R2 Score (c) Pearson's Correlation (d) AUC-ROC Score for the cellular composition prediction task after augmenting limited data with synthetic images.}
\label{data_augmentation_metrics}
\end{figure*}

We also observe changes in appearance of histology images after addition of nuclei or changing their positions. Figure \ref{alter_layout} shows visual results after adding a lymphocyte gets added into the fixed cellular layout and after altering its position. We can also see the appearance of added lymphocyte (in yellow color) in the generated cellular mask. This shows the flexibility of our interactive framework that can be used for customized colon histology image generation.

\begin{table}[]
\resizebox{\columnwidth}{!}{%
\begin{tabular}{|l|c|l|}
\hline
\textbf{Framework} & \textbf{CoNiC} & \textbf{PanNuke} \\ \hline
Pix2Pix                       & 89.35 $\pm$ 0.80  & 107.43 $\pm$ 1.05 \\ \hline
Residual Encoder Decoder      & 86.40 $\pm$ 0.30  & \textbf{102.72 $\pm$ 1.94} \\ \hline
CycleGAN                 & 170.63 $\pm$ 0.97  & 253.30 $\pm$ 2.02 \\ \hline
SynCLay  & \textbf{81.46 $\pm$ 0.25} &  103.48 $\pm$ 1.23 \\ \hline
\end{tabular}%
}
\caption{FID score comparison between our framework and state-of-the-art frameworks for generating histology images conditioned on input nuclei masks}
\label{fid}
\end{table}

\subsection{Assessment by Pathologists}

To assess the realism of synthetic histology images generated using SynCLay framework, we requested 4 pathologists to rate each generated image from 1 (least realistic) to 10 (most realistic). The pathologists were presented with a set of total 30 images from which 15 were real and 15 synthetic. The set contained images generated from different cellular compositions and grades of differentiation. We also included some unnatural images in the set by manipulating nuclei locations such as placing lymphocytes inside glands. 

We requested the pathologists to score each image as well as the associated appearance of nuclei. The scores regarding the quality of the generated tissue images are provided in Table \ref{pathologists_assessment}. We can clearly see that the synthetic images obtain similar average realism score (8.31) compared to that for the real images  (7.92). As can be observed from the Table, average score of synthetic nuclei appearance (8.25) is similar to that of the real nuclei (8.68). Therefore, it can be argued that images generated by the proposed framework appear realistic. However, some pathologists argued that delineation between cells was not clear. Few pathologists found that cytoplasm seemed artificial. Besides, one or two pathologists were able to identify unnatural images generated for the purpose of experiment.


\begin{table}[hbt!]
\centering
\resizebox{\columnwidth}{!}{%
\begin{tabular}{|c|cc|cc|}
\hline
 &
  \multicolumn{2}{c|}{\begin{tabular}[c]{@{}c@{}}Original Data Statistics \\ \#Samples = 3963\end{tabular}} &
  \multicolumn{2}{c|}{\begin{tabular}[c]{@{}c@{}}Balanced Statistics \\ using SynCLay Generated Images\\ \#Samples = 4963\end{tabular}} \\ \hline
Cell Type          & \multicolumn{1}{c|}{Cell Count} & \#Images & \multicolumn{1}{c|}{Cell Count} & \#Images \\ \hline
Neutrophil         & \multicolumn{1}{c|}{3912}       & 1006     & \multicolumn{1}{c|}{20976}      & 2006     \\ \hline
Epithelial         & \multicolumn{1}{c|}{230689}     & 3453     & \multicolumn{1}{c|}{253379}     & 4453     \\ \hline
Lymphocytes        & \multicolumn{1}{c|}{92986}      & 3578     & \multicolumn{1}{c|}{114751}     & 4578     \\ \hline
Plasma             & \multicolumn{1}{c|}{25919}      & 2678     & \multicolumn{1}{c|}{42996}      & 3678     \\ \hline
Eosinophil         & \multicolumn{1}{c|}{3290}       & 1208     & \multicolumn{1}{c|}{22806}      & 2208     \\ \hline
Connective Tissues & \multicolumn{1}{c|}{101643}     & 3664     & \multicolumn{1}{c|}{116377}     & 4664     \\ \hline
\end{tabular}%
}
\caption{Distribution of CoNiC train set. It shows the total cell counts of distinct cell types and also number of images having occurrence of those cells.}
\label{dataaugmentation}
\end{table}

\begin{figure*}[hbt!]
\centering
\includegraphics[width=\textwidth,height=\textheight,keepaspectratio]{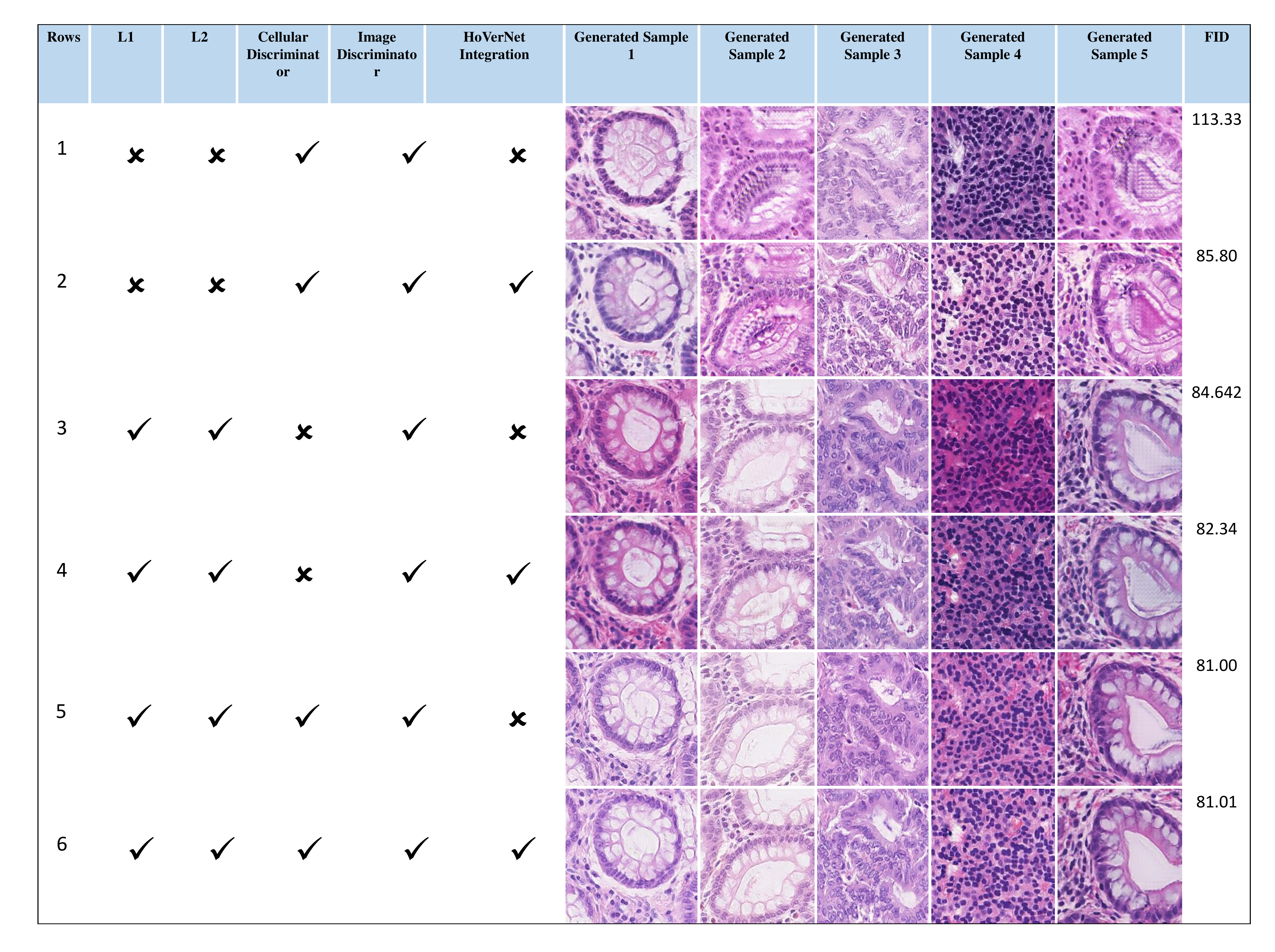}
\caption{Ablation study showing importance of various loss components used in training the proposed framework. Bottom most row shows the generated tissue images with using all of the available components. Top rows show synthetic images generation by keeping some of the total lost components.}
\label{ablation_study}
\end{figure*}

\subsection{Quantitative Analysis}
\label{fid_quality_section}

The Frechet Inception Distance (FID) (\cite{heusel2017gans}) is a widely used metric to evaluate the quality of generated images \citep{quiros2019pathology, Levine2020SynthesisOD, safron} as it quantifies the network's ability to reproduce original data distribution. We assessed the quantitative performance of visual similarity between real tissue images and the corresponding synthetic images generated using the SynCLay framework by computing FID between the two sets of images. 

Table \ref{fid} shows FID scores between a set of real and generated tiles computed by both our framework and other state-of-the-art (SOTA) models on the CoNiC and PanNuke datasets respectively. The lowest FID score between real and synthetic images for the CoNiC dataset suggest that the convolutional feature maps computed from the synthetic images are close to the ones obtained for real images. Though FID score for the SynCLay framework is slightly lower than for the residual encoder-decoder model, it needs to be considered that the SynCLay images are generated from the cellular layouts instead of nuclei masks. Crafting masks can be more cumbersome than drawing up cellular layouts as latter do not require drawing realistic shapes of respective nuclei. Overall, our results suggest that synthetic images are close to realistic images and can potentially be used in computational pathology applications.

\subsection{Synthetic Images for Performance Improvement of Cellular Composition Prediction}

\begin{figure*}[hbt!]
\centering
\includegraphics[width=400pt, keepaspectratio]{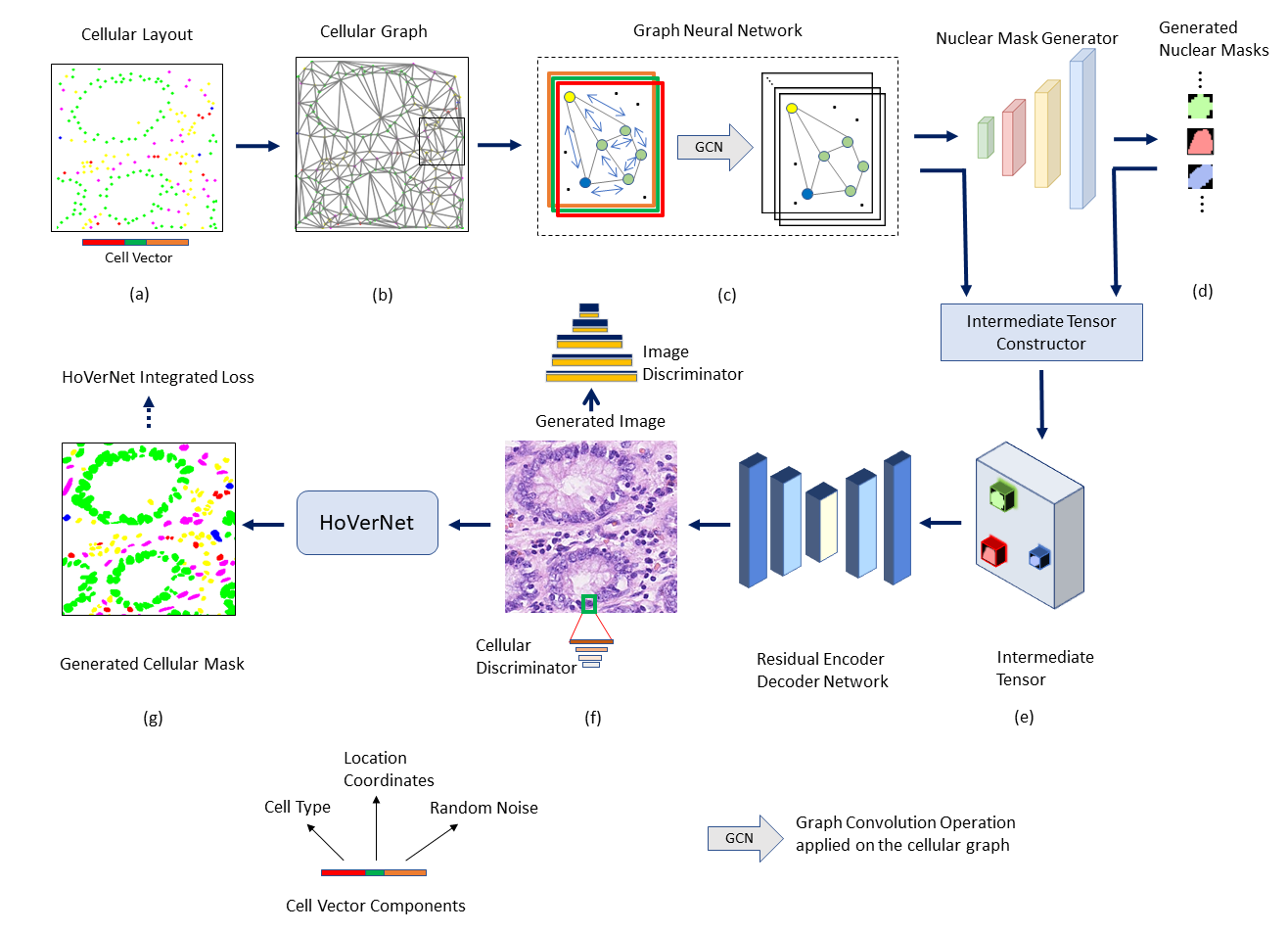}
\caption{SynCLay with Graph Convolution Network. The architecture is very similar to the one shown in \ref{concept_figure}. The only change is a construction of cellular graph (b) and processing it with the graph convolution network (c).}
\label{arch_gcn}
\end{figure*}

The cellular composition prediction refers to the task of evaluating the presence and counts of different types of cells in the tumour microenvironment of Haematoxylin and Eosin (H\&E) stained histology images. The cellular composition analysis can be useful for various downstream prediction tasks in CPath such as survival prediction \citep{Shaban2019AND, survivalprediction2}, gene expression and biological process \citep{geneexpression} and recurrence prediction \citep{recurrenceprediction}. 

The CoNiC dataset is highly unbalanced with respect to counts of different types of cells, as shown in Table \ref{dataaugmentation}. From the table, it can be noticed that neutrophils, and eosinophils are present in a small number of tissue images, additionally their overall incidence in the dataset is less than 1\%. This may affect the performance of cellular composition prediction and nuclei presence detection tasks. In this section, we evaluate the applicability of the proposed SynCLay model to boost the performance of these tasks using a modified version of the ALBRT \citep{albrt} model, which uses five branches (All, Left, Bottom, Right and Top) for predicting the counts of different types of cells in an input image and in their corresponding left, right, top, and bottom halves. We trained ALL branch of ALBRT model for predicting the counts of Neutrophil, Epithelial, Lymphocytes, Plasma, Eosinophil and Connective cells present in a patch. Originally, ALBRT was trained using pairwise ranking loss. However, for this experiment we used Huber loss.
For baseline results, we trained ALBRT model on CoNiC training dataset and assessed model performance on CoNiC validation set. We then balanced the distribution of cellular counts by generating synthetic colon images given the cellularites of different cell types using the proposed SynCLay model. Table \ref{dataaugmentation} shows the distribution of counts of different types of cells in the training dataset used for training ALBRT before and after data augmentation.  We assessed the performance of ALBRT model using both real and synthetic data using the same experimental protocol used for getting baseline results. 

We evaluated the predictive performance of both the models in detecting the presence of different types of cells in a patch using AUC-ROC as performance metric, while for cellular counts prediction we reported Pearson's correlation coefficient, Spearman's correlation coefficient and R2 score between ground truth and predicted cellular counts. The performance comparison of both the models for cellular composition prediction task, and cell presence detection task  can be seen in Figure \ref{data_augmentation_metrics}.

From Figure \ref{data_augmentation_metrics}, it can be seen that cellular composition and cell presence detection performance of the ALBRT model improved significantly for neutrophils and eosinophils after addition of cell-type specific synthetic images in the training data generated using the proposed framework. For example, the AUC-ROC improved by 4\% and 6\% for neutrophils and eosinophils, respectively, when using synthetic images in training. Spearman's correlation coefficient registered an increase of 11\% for neutrophils, and 8\% for eosinophils. Similarly, Pearson's correlation coefficient demonstrated the gain of 7\% for neutrophils and 5\% for eosinophils. The improvement in performance validates the utility of customized synthetic images for the task of cellular composition prediction. The performance metrics for rest of the cells including epithelial cells, lymphocytes, plasma cells and connective tissues look nearly equal. This might be due to the fact that the cellular counts and number of images containing those cells were relatively high and sufficient already for training the ALBRT model.

\subsection{Ablation Studies}
\label{ablationstudy}

In this section, we perform an ablation study to examine the importance of each of the loss terms used to train the SynCLay framework. We also assess and compare the performance of SynCLay after incorporation of the graph neural network for processing cellular vectors to generate histology images. 
\begin{figure*}[hbt!]
\centering
\includegraphics[width=\textwidth,height=\textheight,keepaspectratio]{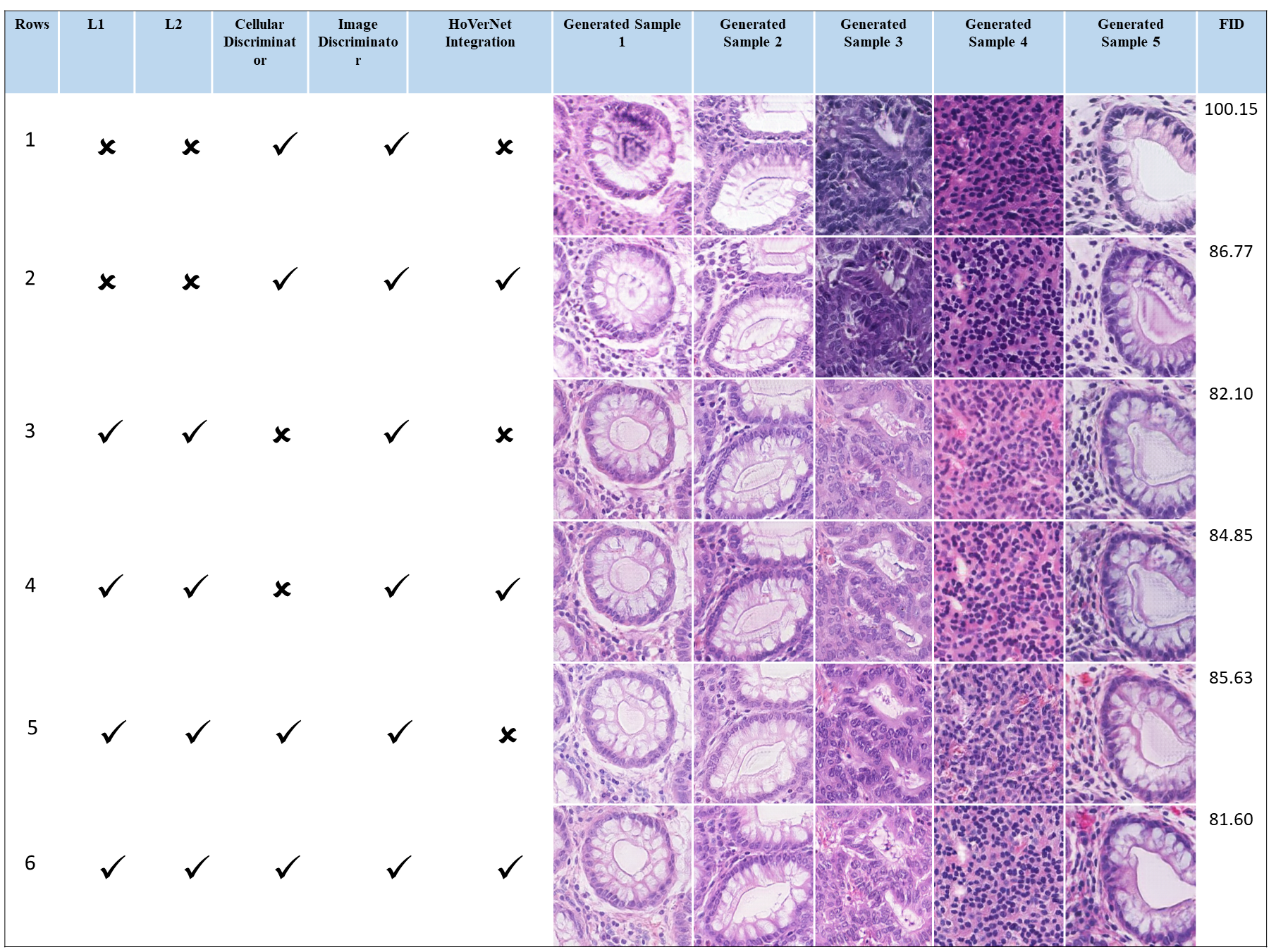}
\caption{Ablation study showing importance of various loss components used in training the proposed SynCLay  framework with the graph convolution network. Bottom most row shows the generated tissue images with using all of the available components. Top rows show synthetic images generation by keeping some of the total lost components.}
\label{ablation_study_gcn}
\end{figure*}

\subsubsection{Assessments of Loss Terms}

We train the SynCLay  network by using different combinations of the loss terms discussed in section \ref{loss_components}. The visual and quantitative results of this experiment are presented in Table \ref{ablation_study}. 

It can be observed that the generated images exhibit major and minor patch artifacts after removal of L1 \& L2 loss and by detaching the cellular discriminator respectively, specifically in first four rows in Table \ref{ablation_study}. We can also notice that enabling HoverNet integrated loss moderately increased the fidelity of generated images in terms of FID, for instance, FID reduction from 113.33 to 85.80 in first two rows and from 84.64 to 82.32 in next two rows in Table \ref{ablation_study}. It is interesting to notice that the model trained with all of the loss terms shows almost equal FID with or without HoVer-Net but enabling HoVer-Net improves the visual quality of synthetic images. Specifically, the nuclei structures and lumen components inside the tissue images.

\subsubsection{Graph Convolution Operation}

Methods by \citep{scene_graph, scene_graph_2} generate natural images from the input scene graphs. They employ a graph convolution operation to process objects in the scene graphs along with their inter-dependencies. It is obvious that the graph convolution operation can be employed to process cellular vectors of cells inside the cellular layout. We incorporated the graph convolution operation in the SynCLay architecture shown in Figure \ref{concept_figure}. The modified architecture can be seen in Figure \ref{arch_gcn}. Specifically, we construct the cellular graph by computing Delaunay Triangulation \citep{delaunay} over 2-dimensional cellular locations. The edges between cells are evaluated by computing Euclidean distances between spatial locations. Cellular graph is processed by a \textit{graph convolution network} (GCN) \citep{graph_neural, scene_graph} that generates the per-cell latent embedding. These latent embeddings now act as the cellular vectors while rest of the architecture remains the same as described in the section \ref{materials_methods}.

We train the SynCLay framework with GCN on the CoNiC train set with same settings used while training SynCLay without GCN with varied loss terms as described in the previous section, and plot the results in Table \ref{ablation_study_gcn} akin to in Table \ref{ablation_study}. From both the tables, it is apparent that adding GCN adds little improvement as both cases are showing similar visual and quantitative measures. A potential reason can be that constructing the nuclei shapes is relatively less complicated compared to that of objects in natural images \citep{scene_graph, scene_graph_2}. Therefore, using SynCLay without graph convolution network can be thought as an advantage in the CPath domain as the architecture becomes relatively computationally and memory efficient compared to the one with graph convolution network.

\section{Discussion}
\label{discussion}

A limitation of our method is that it requires cellular layouts to generate high-quality colon histology images. However, we have shown how layouts can be created using the THeCoT method (\cite{kovacheva2016model}). We have also presented utility of these masks and generated synthetic images in augmenting limited data for the cellular composition prediction algorithm. After combination of the proposed framework with TheCoT model enables the generation of histology images from a set of user-defined parameters such as grade of differentiation, cellularities of different cells.

The proposed SynCLay framework can be thought of as a crucial step towards construction of complete whole-slide images from a set of user-defined parameters. Whole-slide images are generally multi-gigapixel in nature and have great mixture of various tissue components with large number of inter-dependencies among them. The object layouts where tissue components can be placed on the Cartesian plane can be generated or acquired using segmentation techniques depending on the prediction task or using frameworks similar to TheCoT. The proposed framework can be adapted to consume object layouts to construct the whole-slide images. The framework can potentially be useful in generating an exhaustive annotated data for training and evaluation of tasks like cellular composition prediction and nuclei segmentation in the domain of computational histopathology. Moreover, the discriminator output can be utilized to compute importance weights as given in (\cite{robustlabelsynthesize}). The weighted training can potentially increase the performance of nuclei segmentation algorithms.

\section{Conclusions \& Future Directions}

We presented a novel interactive framework SynCLay for generation of synthetic colon histology image tiles from cellular layouts. We showed that the synthetic tissue images generated by the framework appear realistic and preserve morphological characteristics in the tissue regions. We also assessed quality of the generated tissue images using the FID metric. Synthetic images generated using our framework showed consistently low FID and outperformed those generated by other SOTA models by a significant margin. We assessed the quality of generated images with the help of pathologists and found that synthetic images appear highly realistic from the pathologist's point of view. We demonstrated that the synthetic image tiles constructed from our framework, accompanied with a definitive ground-truth generated by a parametric model, can be used for development of deep learning algorithms for computational histopathology tasks like cellular composition prediction, especially when we have highly limited and unbalanced data.

The proposed framework may be used to extend already existing segmentation datasets for histology image analysis, enabling researchers to improve the performance of automated segmentation approaches for computational pathology. This framework can be generalized for producing a large number of customized images for different types of carcinomas and tissue types. An open future direction for research is to extend the proposed framework to generate complete whole-slide images from known parameters.

\section*{Acknowledgements}
Fayyaz Minhas and Nasir Rajpoot are part of the PathLAKE digital pathology consortium, which is funded from the Data to Early Diagnosis and Precision Medicine strand of the governments Industrial Strategy Challenge Fund, managed and delivered by UK Research and Innovation (UKRI). Nasir Rajpoot was also supported by the UK Medical Research Council (grant award MR/P015476/1) and the Alan Turing Institute. Muhammad Dawood is co-funded by GSK and the department of computer science at University of Warwick.

We thank collaborating pathologists for their valuable input in assessing images generated using the proposed method. In particular, we are grateful to Drs David Snead, Robinson Andrew, Hesham ElDaly and Yee-Wah Tsang.

\clearpage 

\bibliographystyle{model2-names.bst}\biboptions{authoryear}
\bibliography{refs}

\newpage 

\section*{Appendix}
\label{appendix}

Here we describe all network architectures for all components in our proposed framework. 

\subsection*{Mask Generator Network}

We generate the binary nuclei mask for each of the cells using the \textit{mask generator network}. The input is the individual latent vectors obtained after applying the \textit{graph convolution operation} on the cellular graphs, and output is the $64 \times 64$ cellular mask ($M=64$) with all elements ranged between 0 and 1. The mask generator network composed of series of mask generator blocks, where each block consist of interpolation + convolution + batch normalization + ReLU activation operations. The exact architecture is shown in the table \ref{tabletwo}, while architecture of the \textit{mask generator block} is shown in Table \ref{three}.

\begin{table}[hbt!]
\centering
\begin{tabular}{|c|c|c|c|}
\hline
\textbf{Index} & \textbf{Inputs} & \textbf{Operation}                            & \textbf{Output Shape} \\ \hline
(1)            & -               & Latent Embedding                        & 32                    \\ \hline
(2)            & (1)             & Reshape                                       & 32 x 1 x 1            \\ \hline
(3)            & (2)             & Mask Generator Block                          & 32 x 2 x 2            \\ \hline
(4)            & (3)             & Mask Generator Block                          & 32 x 4 x 4            \\ \hline
(5)            & (4)             & Mask Generator Block                          & 32 x 8 x 8            \\ \hline
(6)            & (5)             & Mask Generator Block                          & 32 x 16 x 16          \\ \hline
(7)            & (6)             & Mask Generator Block                          & 32 x 32 x 32          \\ \hline
(8)            & (7)             & Mask Generator Block                          & 32 x 64 x 64          \\ \hline
(9)            & (8)             & Conv2d (K=1, 32 $\rightarrow$ 1) & 1 x 64 x 64           \\ \hline
(10)            & (9)             &  Sigmoid & 1 x 64 x 64           \\ \hline
\end{tabular}
\caption{Architecture of the mask generator network. The function implements function $M$ from the main text.  The notation Conv2d(K , $C_{in}$ → $C_{out}$) is a convolution with $K \times K$ kernels, $C_{in}$ input channels and $C_{out}$ output channels; all convolutions with stride 1 with zero padding that ensures input and output have the same spatial size.}
\label{tabletwo}
\end{table}

\begin{table}[hbt!]
\centering
\begin{tabular}{|c|c|}
\hline
\textbf{Operation}                             & \textbf{Output Shape} \\ \hline
Interpolation                                  & 32 x 2S x 2S          \\ \hline
Conv2d (K=3, 32 $\rightarrow$ 32) & 32 x 2S x 2S          \\ \hline
Batch Normalization                            & 32 x 2S x 2S          \\ \hline
ReLU                                           & 32 x 2S x 2S          \\ \hline
\end{tabular}
\caption{Architecture of the mask generator block. The input is the feature map of shape $C \times S \times S$, where C is the number of channels from the feature map of the last layer, and $S \times S$ is the dimension of height and width.}
\label{three}
\end{table}

\begin{table*}[hbt!]
\centering
\begin{tabular}{|c|c|c|c|}
\hline
\textbf{Index} & \textbf{Inputs} & \textbf{Operation}                             & \textbf{Output Shape} \\ \hline
(1)            & -               & Generate Cumulative Mask                        & 32 x 256 x 256        \\ \hline
(2)            & (1)             & Conv2d (K=3, 32 $\rightarrow$ 16) & 16 x 256 x 256        \\ \hline
(3)            & (2)             & LeakyReLU                                      & 16 x 256 x 256        \\ \hline
(4)            & (3)             & Conv2d (K=3, 16 $\rightarrow$ 8)  & 8 x 256 x 256         \\ \hline
(5)            & (4)             & LeakyReLU                                      & 8 x 256 x 256         \\ \hline
(6)            & (5)             & Conv2d (K=3, 8 $\rightarrow$ 4)   & 4 x 256 x 256         \\ \hline
(7)            & (6)             & LeakyReLU                                      & 4 x 256 x 256         \\ \hline
(8)            & (7)             & Conv2d (K=3, 8 $\rightarrow$ 4)   & 1 x 256 x 256         \\ \hline
(9)            & (8)             & LeakyReLU                                      & 1 x 256 x 256         \\ \hline
\end{tabular}
\label{channelreducer}
\caption{Architecture of the channel reducer network. The network implements function $R$ from the main text. LeakyReLU uses a negative slope coefficient of 0.2}
\end{table*}

\subsection*{Encoder-Decoder Network}

The final tissue image is generated from the generated tissue component mask with the help of \textit{encoder decoder network}. The input is the intermediate tensor generated using the bilinear interpolation operation on latent embeddings and generated binary per nuclei masks. The encoder consist of a series of ``Encode" blocks (shown in Table \ref{encoderblock}) and generates the lower sized encoding of the input mask, while decoder comprised of a series of ``Decode" blocks (shown in Table \ref{decodeblock}) and generates the final tissue image from the encoding. The exact architecture of the encoder-decoder network is shown in Table \ref{encoderdecodernetwork}. 

\begin{table}[hbt!]
\centering
\begin{tabular}{|c|c|c|c|}
\hline
\textbf{Index} & \textbf{Inputs} & \textbf{Operation}                             & \textbf{Output Shape} \\ \hline
(1)            & -               & Intermediate Tensor                        & 32 x 256 x 256         \\ \hline
(2)            & (1)             & Encode(1,64)                                   & 64 x 128 x 128        \\ \hline
(3)            & (2)             & Encode(64,128)                                 & 128 x 64 x 64         \\ \hline
(4)            & (3)             & Encode(128,256)                                & 256 x 32 x 32         \\ \hline
(5)            & (4)             & Encode(256,512)                                & 512 x 16 x 16         \\ \hline
(6)            & (5)             & Encode(512,512)                                & 512 x 8 x 8           \\ \hline
(7)            & (6)             & Encode(512,512)                                & 512 x 4 x 4           \\ \hline
(8)            & (7)             & Encode(512,512)                                & 512 x 2 x 2           \\ \hline
(9)            & (8)             & Encode(512,512)                                & 512 x 1 x 1           \\ \hline
(10)           & (9,8)           & Decode(512,512)                                & 1024 x 2 x 2          \\ \hline
(11)           & (10,7)          & Decode(1024,512)                               & 1024 x 4 x 4          \\ \hline
(12)           & (11,6)          & Decode(1024,512)                               & 1024 x 8 x 8          \\ \hline
(13)           & (12,5)          & Decode(1024,512)                               & 1024 x 16 x 16        \\ \hline
(14)           & (12,4)          & Decode(1024,256)                               & 512 x 32 x 32         \\ \hline
(15)           & (14,3)          & Decode(512,128)                                & 256 x 64 x 64         \\ \hline
(16)           & (15,2)          & Decode(256,64)                                 & 128 x 128 x 128       \\ \hline
(17)           & (16)            & Upsample                                       & 128 x 256 x 256       \\ \hline
(18)           & (17)            & Conv2d (K=4, 128 $\rightarrow$ 3) & 3 x 256 x 256         \\ \hline
(19)           & (18)            & Tanh                                           & 3 x 256 x 256         \\ \hline
\end{tabular}
\caption{Architecture of the encoder-decoder residual network. The network implements the function $E$ from the main text.}
\label{encoderdecodernetwork}
\end{table}

\begin{table}[hbt!]
\centering
\begin{tabular}{|c|c|}
\hline
\textbf{Operation}                                & \textbf{Output Shape} \\ \hline
Conv2d (K=4, $C_{in} \rightarrow C_{out})$ & $C_{out}$ x S x S          \\ \hline
Instance Normalization (if normalize=True)        & $C_{out}$ x S x S          \\ \hline
LeakyReLU                                         & $C_{out}$ x S x S          \\ \hline
Dropout (if dropout=True)                         & $C_{out}$ x S x S          \\ \hline
\end{tabular}
\caption{Architecture of the ``Encode" block. LeakyReLU uses a negative slope coefficient of 0.2}
\label{encoderblock}
\end{table}

\begin{table}[hbt!]
\centering
\begin{tabular}{|c|c|}
\hline
\textbf{Operation}                                        & \textbf{Output Shape} \\ \hline
ConvTranspose2d(K=4, $C_{in} \rightarrow C_{out}$) & $C_{out}$ x S x S          \\ \hline
Instance Normalization                                    & $C_{out}$ x S x S          \\ \hline
ReLU                                                      & $C_{out}$ x S x S          \\ \hline
Dropout (if dropout=True)                                 & $C_{out}$ x S x S          \\ \hline
\end{tabular}
\caption{Architecture of the ``Decode" block}
\label{decodeblock}
\end{table}

\subsection*{Image Discriminator}

The discriminator we employed for tissue images ($D_I$), takes the real or fake image of shape $3 \times 256 \times 256$ as an input and classifies an overlapping grid of size $7 \times 7$ image patches from the input image as real or fake. The exact architecture of the discriminator is shown in Table \ref{discriminatornetwork}

\subsection*{Cellular Discriminator}

The cellular discriminator $D_C$ consumes image pixels corresponding to cellular areas from the real or fake tissue images, and classifies them as real or fake. The cellular areas are cropped out using their bounding box coordinates, and resized to $64 \times 64$ pixels using the bilinear interpolation method. The exact architecture of the cellular discriminator is shown in Table \ref{celldiscriminatornetwork}

\begin{table}[]
\centering
\begin{tabular}{|c|c|c|c|}
\hline
\textbf{Index} & \textbf{Inputs} & \textbf{Operation}                                  & \textbf{Output Shape} \\ \hline
(1)            & -               & Crop cellular portions from the generated image    & 3 x 64 x 64           \\ \hline
(2)            & (1)             & Conv2d (K=5, 3 $\rightarrow$ 16, S=2)  & 16 x 30 x 30          \\ \hline
(3)            & (2)             & Batch Normalization                                 & 16 x 30 x 30          \\ \hline
(4)            & (3)             & LeakyReLU                                           & 16 x 30 x 30          \\ \hline
(5)            & (4)             & Conv2d (K=5, 16 $\rightarrow$ 32, S=2) & 32 x 13 x 13          \\ \hline
(6)            & (5)             & Batch Normalization                                 & 32 x 13 x 13          \\ \hline
(7)            & (6)             & LeakyReLU                                           & 32 x 13 x 13          \\ \hline
(8)            & (7)             & Conv2d (K=5, 32 $\rightarrow$ 64, S=2) & 64 x 5 x 5            \\ \hline
(9)            & (8)             & Global Average Pooling                              & 64                    \\ \hline
(10)           & (9)             & Affine Transformation                               & 1024                  \\ \hline
(11)           & (10)            & Affine Transformation                               & 1                     \\ \hline
\end{tabular}
\caption{Architecture of the cellular discriminator, $D_{C}$. LeakyReLU uses a negative slope coefficient of 0.2}
\label{celldiscriminatornetwork}
\end{table}

\begin{table*}[hbt!]
\centering
\begin{tabular}{|c|c|c|c|}
\hline
\textbf{Index} & \textbf{Inputs} & \textbf{Operation}                                    & \textbf{Output Shape} \\ \hline
(1)            & -               & Generate the Image                                    & 3 x 256 x 256         \\ \hline
(2)            & (1)             & Conv2d (K=4, C $\rightarrow$ 16, S=2)    & 16 x 128 x 128        \\ \hline
(3)            & (2)             & LeakyReLU                                             & 16 x 128 x 128        \\ \hline
(4)            & (3)             & Conv2d (K=4, 16 $\rightarrow$ 32, S=2)   & 32 x 64 x 64          \\ \hline
(5)            & (4)             & LeakyReLU                                             & 32 x 64 x 64          \\ \hline
(6)            & (5)             & Instance Normalization                                & 32 x 64 x 64          \\ \hline
(7)            & (6)             & Conv2d (K=4, 32 $\rightarrow$ 64, S=2)   & 64 x 32 x 32          \\ \hline
(8)            & (7)             & LeakyReLU                                             & 64 x 32 x 32          \\ \hline
(9)            & (8)             & Instance Normalization                                & 64 x 32 x 32          \\ \hline
(10)           & (9)             & Conv2d (K=4, 64 $\rightarrow$ 128, S=2)  & 128 x 16 x 16         \\ \hline
(11)           & (10)            & LeakyReLU                                             & 128 x 16 x 16         \\ \hline
(12)           & (11)            & Instance Normalization                                & 128 x 16 x 16         \\ \hline
(13)           & (12)            & Conv2d (K=4, 128 $\rightarrow$ 256, S=2) & 256 x 8 x 8           \\ \hline
(14)           & (13)            & LeakyReLU                                             & 256 x 8 x 8           \\ \hline
(15)           & (14)            & Instance Normalization                                & 256 x 8 x 8           \\ \hline
(16)           & (15)            & Conv2d (K=4, 256 $\rightarrow$ 1 , S=1)  & 1 x 7 x 7             \\ \hline
\end{tabular}
\caption{Architecture of the Discriminator network. All but the last Conv2d operation has stride 2. LeakyReLU uses a negative slope coefficient of 0.2}
\label{discriminatornetwork}
\end{table*}

\newpage

\end{document}